\documentclass[a4paper]{jpconf}
\usepackage{graphicx}
\begin{document}
\title{The MPGD-Based Photon Detectors for the upgrade of 
COMPASS RICH-1}
\author{J.~Agarwala, S.~Dalla~Torre, S.~Dasgupta, M.~Gregori, G.~Hamar, S.~Levorato, G.~Menon, F.~Tessarotto, Triloki, Y.X.~Zhao}
\address{INFN Trieste, Trieste, Italy}

\author{F.~Bradamante, A.~Bressan, C.~Chatterjee, P.~Ciliberti, A.~Martin }
\address{University of Trieste and INFN Trieste, Trieste, Italy}

\author[add-to-infn]{M.~Alexeev, O.~Denisov}
\address{INFN Torino, Torino, Italy}

\author{M.~Chiosso}
\address{University of Torino and INFN Torino, Torino, Italy}

\author{D.~Panzieri}
\address{University of East
Piemonte, Alessandria and INFN Torino, Torino, Italy}

\author{C.D.R.~Azevedo, J.F.C.A.~Veloso}
\address{University of Aveiro, Aveiro, Portugal}

\author{M.~B\"uchele,  H.~Fischer, F.~Herrmann, S. Schopferer }
\address{Universit\"at Freiburg, Freiburg, Germany}

\author{A.~Cicuttin, M.L.~Crespo}
\address{Abdus Salam ICTP, Trieste and INFN Trieste, Trieste, Italy}

\author{M.~Finger, M.~Finger~Jr, M. Slunecka}
\address{Charles
University, Prague, Czech Republic and JINR, Dubna, Russia}

\author{M. Sulc}
\address{Technical University of Liberec, Liberec, Czech Republic}

\ead{silvia.dallatorre@ts.infn.it}

\begin{abstract}
After pioneering gaseous detectors of single photon for RICH 
applications using CsI solid state photocathodes in MWPCs within 
the RD26 collaboration and by the constructions for the RICH detector 
of the COMPASS experiment at CERN SPS, in 2016 we have upgraded 
COMPASS RICH by novel gaseous photon detectors based on MPGD 
technology. Four novel photon detectors, covering 
a total active area 
of 1.5~m$^2$, have been installed in order to cope with the 
challenging efficiency and stability requirements 
of the COMPASS
physics programme. 
They are the first application in an 
experiment of MPGD-based single photon detectors.
All aspects of the upgrade are presented, including engineering, 
mass production, quality assessment and performance.
\par
Perspectives for further developments in the field of gaseous 
single photon detectors are also indicated.
\end{abstract}
\section{Introduction}
THE RICH-1 detector~\cite{rich-1} of the 
COMPASS Experiment~\cite{compass}
at CERN SPS has been
upgraded: four new Photon Detectors 
(unit size: 600$\times$600~mm$^2$),
based on MPGD technology and covering a total active area
of 1.5~m$^2$ replace the previously used MWPC-based photon
detectors in order to cope with the challenging efficiency and
stability requirements of the new COMPASS measurements.
In fact,
COMPASS goal is to deal with trigger rates up to O(10$^5$)~Hz
and beam rates up to O(10$^8$)~Hz. Concerning increased 
stability, this is provided by the new detector 
architecture, as explained in the following.
In COMPASS RICH-1, MPGD photon detectors are used for
the first time in a running experiment. This realization 
also opens the way of a more extended
use of novel gaseous photon detectors in the domain of the
Cherenkov imaging technique for Particle IDentification (PID),
key detectors in several research sectors and, in particular, in
hadron physics. The relevance is related to the role of gaseous
photon detectors, which are still the only available option to
instrument detection surfaces when insensitivity to magnetic
field, low material budget, and affordable costs in view of
large detection systems are required. The MPGD-based photon
detectors overcome the limitation of the previous generation of
gaseous photon detectors thanks to two essential performance
characteristics: reduced ion and photon backflow to the photocathode,
namely reduced ageing and increased electrical stability, and
faster signal development, namely higher rate capabilities.
\section{The novel photon detectors}
\begin{figure}
\centering
\includegraphics[width=0.99\linewidth]{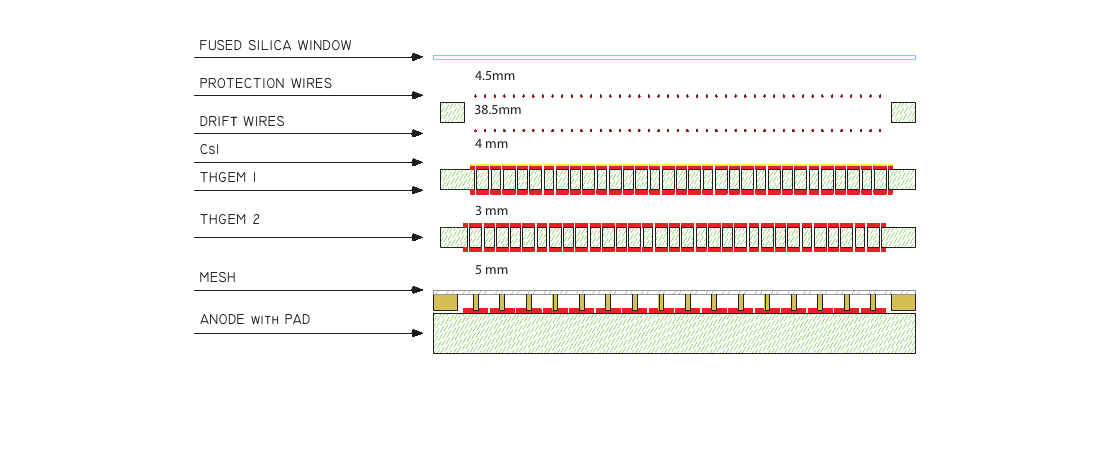}
\caption{Sketch of the hybrid single photon detector: 
two staggered THGEM
layers are coupled to a resistive bulk MM. Image not to scale.}
\label{fig:architecture}
\end{figure}

The detector architecture is the result of a 
seven-year R\&D activity~\cite{rd}. It is based on a  hybrid
MPGD combination (Fig.~\ref{fig:architecture}), consisting in
two layers of THick GEMs
(THGEM)~\cite{thgem} followed by a resistive 
MicroMegas (MM)~\cite{mm} on a pad segmented
anode. The first THGEM 
also acts as a reflective
photocathode: its top face is coated with a CsI film.
The feedback of photons generated in the multiplication 
process is suppressed by the presence of two THGEM layers, 
while the large majority of the ions from multiplication
are trapped in the MM stage. MPGD properties ensure
signal development in about 100 ns. 
\par 
Each of the four large  (600$\times$600~mm$^2$) single photon
detectors is formed by two identical modules 600$\times$300~mm$^2$,
arranged side by side. The THGEM geometrical parameters are: 
470~$\mu$m thickness,
400~$\mu$m hole diameter and 800~$\mu$m pitch. Holes
are rim-less, 
namely there is no uncoated area
around the hole edge. 
They are arranged in a regular pattern with equilateral triangles 
as elementary cell.
In order to mitigate the effect of 
occasional discharges, the top and bottom electrodes 
of each THGEM are
segmented in 12 parallel areas separated by 0.7~mm clearance, 
each biased via an individual protection 500~M$\Omega$ 
resistor. Therefore, 
discharges only affect a single sector and the operating conditions 
are restored in about 10~s.
The two layers are 
staggered, namely there is complete misalignment between 
the two set of holes: it is so possible to enlarge the electron 
cloud reaching the MM stage, therefore favoring larger gain
in the last amplification stage.
\par
\begin{figure}
\centering
\includegraphics[width=0.7\linewidth]{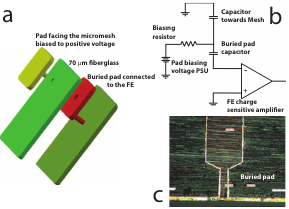}
\caption{a) Sketch of the capacitive coupled readout pad. 
The biasing voltage is
distributed via independent 470~M$\Omega$
resistors to the pad facing the micromesh
structure
(yellow pad in the sketch). The buried pad (red pad in the sketch)
is isolated via 70~$\mu$m 
thick fiberglass and connected
to the front end chip. 
b) Schematic of the capacitive coupled pad principle
illustrated via discrete element blocks. 
c) Metallography section of the PCB:
detail of the through-via connecting the 
external pad through the hole of the
buried pad. The reduced diameter of the through-via
reaching the external pad contributes preserving the pad planarity.}
\label{fig:resistive-mm}
\end{figure}

The MMs have a 
gap of 128~$\mu$m;
they are built by the MM bulk technology~\cite{bulk} 
using 300~$\mu$m
diameter pillars with 2~mm pitch. 
The
MM anode is segmented in 7.5$\times$7.5~mm$^2$ pads. 
The MM resistivity
is realized through 
an original implementation, where no resistive layer is applied
to the pads: the resistivity is obtained by 470~M$\Omega$ resistors in
series with each individual pad
(Fig.~\ref{fig:resistive-mm}). 
The 0.5~mm clearance
between pads prevents the occasional discharges to
propagate towards the surrounding pads: the voltage drop of
the anodic pads surrounding a tripping one is about 2 V over
the typical 600 V operation voltage, 
causing a local gain drop lower
than 4\%. The nominal voltage condition of the pad where 
the discharge occurred
is restored in about 1~s.
The detector is operated with
Ar:CH$_4$~=~50:50 gas mixture, which ensures effective 
extraction of photoelectrons from the photocathode.   
The typical voltage applied are 1270~V across THGEM1, 1250~V 
across THGEM2, and
620~V to bias the MM. 
The drift field above the first THGEM is 
500~V/cm, the transfer field between the two THGEMs is 1000~V/cm
and the field between the second THGEM and the MM micromesh 
is 1000~V/cm.
The effective gain-values
for the three multiplication layers are 
around 12, 10 and 120; these values include the electron transfer 
efficiency.
\par
The novel detectors are read out by the
read-out system already used for the MWPCs with CsI photocathode.
This read-out system is based on the APV front-end chip~\cite{apv}
read out by a dedicated ADC~\cite{apv-system}.
\section{Construction, quality control of the components, assembly and installation}
The electrical stability of large-size THGEMs is a critical issue.
A dedicated protocol  has been elaborated for finishing the industrial produced THGEMs~\cite{shudda-thesis}. It includes polishing 
with fine grain pumice powder, cleaning with water at high pressure, ultrasonic bath with
Sonica PCB solution (PH11), rinsing with distilled water
and backing in oven at 160$^o$C. The procedure moves 
THGEM breakdown voltage to at least 90\% of 
the phenomenological
Paschen limit~\cite{paschen}.
The quality control of the detector components includes:
\begin{itemize} 
\item
the preselection of the raw material for the PCB that will
form the THGEMs in order to use only foils with
homogeneous thickness to guarantee the homogeneity of
the gain;
\item 
the THGEM control by optical inspection, by collecting and
analyzing microscope images, scanning by samples the
large multiplier surface;
\item
the THGEM validation by gain maps using the multipliers in
single layer detectors; gain uniformity at 7\% r.m.s. is obtained;
\item
the collection of MM gain maps illuminating the
detectors by an X-ray gun station; gain uniformity 
at 5\% r.m.s. is obtained;
\item
the measurement of the quantum efficiency of the CsI 
photocathodes, which is performed immediately after 
the coating process; the uniformity within a photocathode is at 
the 3\% level r.m.s. and among the whole production at 
the 10\% level r.m.s.;
\item
the gas leak checks and overall electrical stability checks of
the final detectors.
\end{itemize}
CsI photocathodes must never be 
exposed to air to fully preserve their quantum efficiency: in fact,
CsI is highly hygroscopic and it reacts with water vapour 
that decomposes the molecule. Therefore, 
the presence of CsI photocathodes 
imposes to perform
detector assembly,  transportation and
installation in  glove boxes flushed with N$_2$. 
\section{The high voltage system}
An essential tool for the detector commissioning is the High
Voltage (HV) control system, which also allows for voltage and
current monitoring and data logging. The power supplies are
commercial ones by CAEN inserted in a SY4527 mainframe. 
A1561HDN  units power the THGEMs, while A7030DP power supplies are 
used for the MMs.
The four detectors are organized, from the HV supply point of
view, in four independent sectors each; nine different electrode
types, each one with its specific role, are present in the
multilayer detectors. The total number of HV channels is 136. 
Manual setting and control of all these
HV channels would be both unpractical and unsafe.
The voltages and currents of all the channels are read-out
and recorded at 1~Hz frequency. If the current spark rate is
above a given value the voltage is automatically readjusted.
The system also provides automatic voltage adjustment to 
compensate for the variation of the environmental parameters, 
namely pressure and temperature, that can affect the detector gain. 
Gain stability at 
the 5\% level over months of operation has been obtained.
\section{Preliminary performance results}
The novel detectors have been used during COMPASS runs in year 2016 
and 2017, for a total running period of about 12 months
at COMPASS nominal beam rates.
No HV trip is observed during detector operation: 
thanks to the resistors protecting the THGEM segments and 
the MM pads, in case of occasional discharges, 
only current sparks 
are observed, which temporary affect the local performance. 
The sparks 
in the two THGEM layers are fully correlated. The sparks 
observed in the MM are induced by the THGEM sparks. The 
restoration after a current spark 
is completed within 10~s and the current spark rate is 
typically 1/h/detector (600$\times$600~mm$^2$). These figures result in totally negligible dead-time related to sparks.
\par
The measured rate of ion backflow to the photocathode is at the 3\% level.
The electronics noise, substantially uniform over the detector surface, is 
at the 900 electrons equivalent level (r.m.s).
\par
\begin{figure}
\centering
\includegraphics[width=0.8\linewidth]{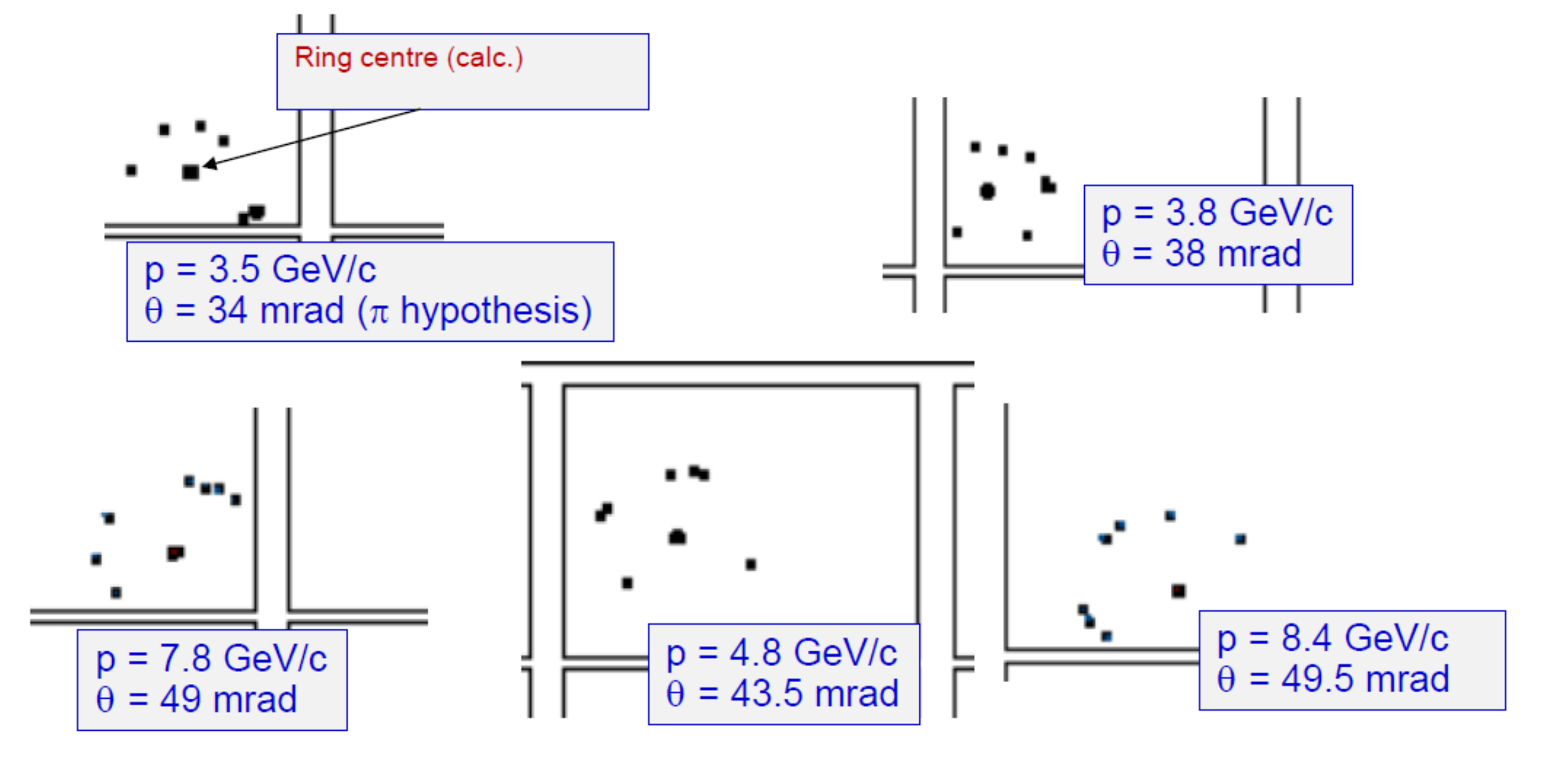}
\caption{Images of hit pattern in the novel photon detectors.
The center of the expected ring patterns is obtained 
from the reconstructed particle trajectories; the particle momentum 
and the expected Cherenkov angle in the pion hypothesis are also reported. 
No image elaboration or background subtraction is applied.}
\label{fig:rings}
\end{figure}
\begin{figure}
\centering
\includegraphics[width=0.8\linewidth]{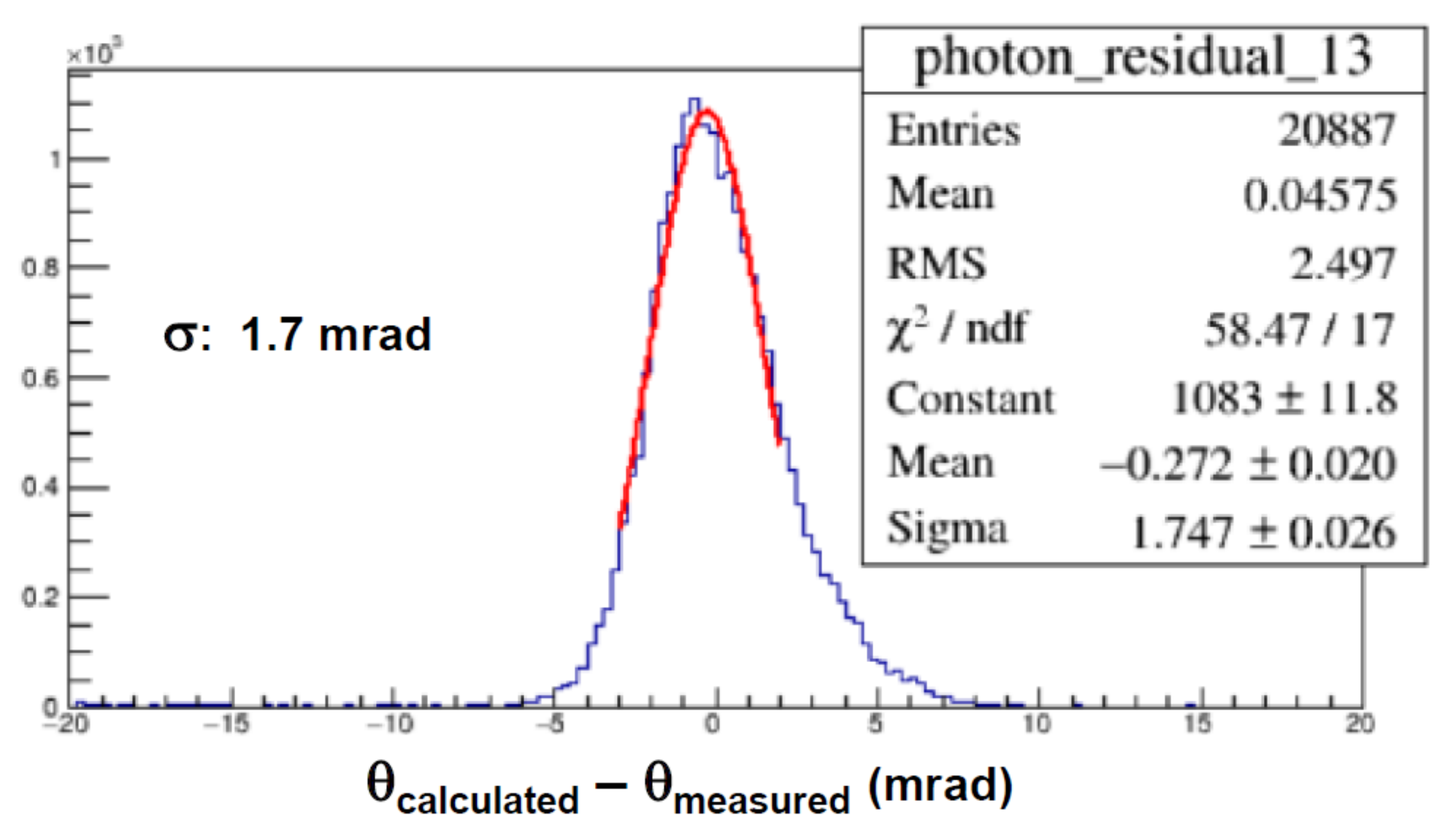}
\caption{Distribution of the difference between the Cherenkov angle
calculated from the reconstructed particle momentum and the 
Cherenkov angle provided by single detected photoelectrons;  a sample of identified pions is used.}
\label{fig:resolution}
\end{figure}
\begin{figure}
\centering
\includegraphics[width=0.8\linewidth]{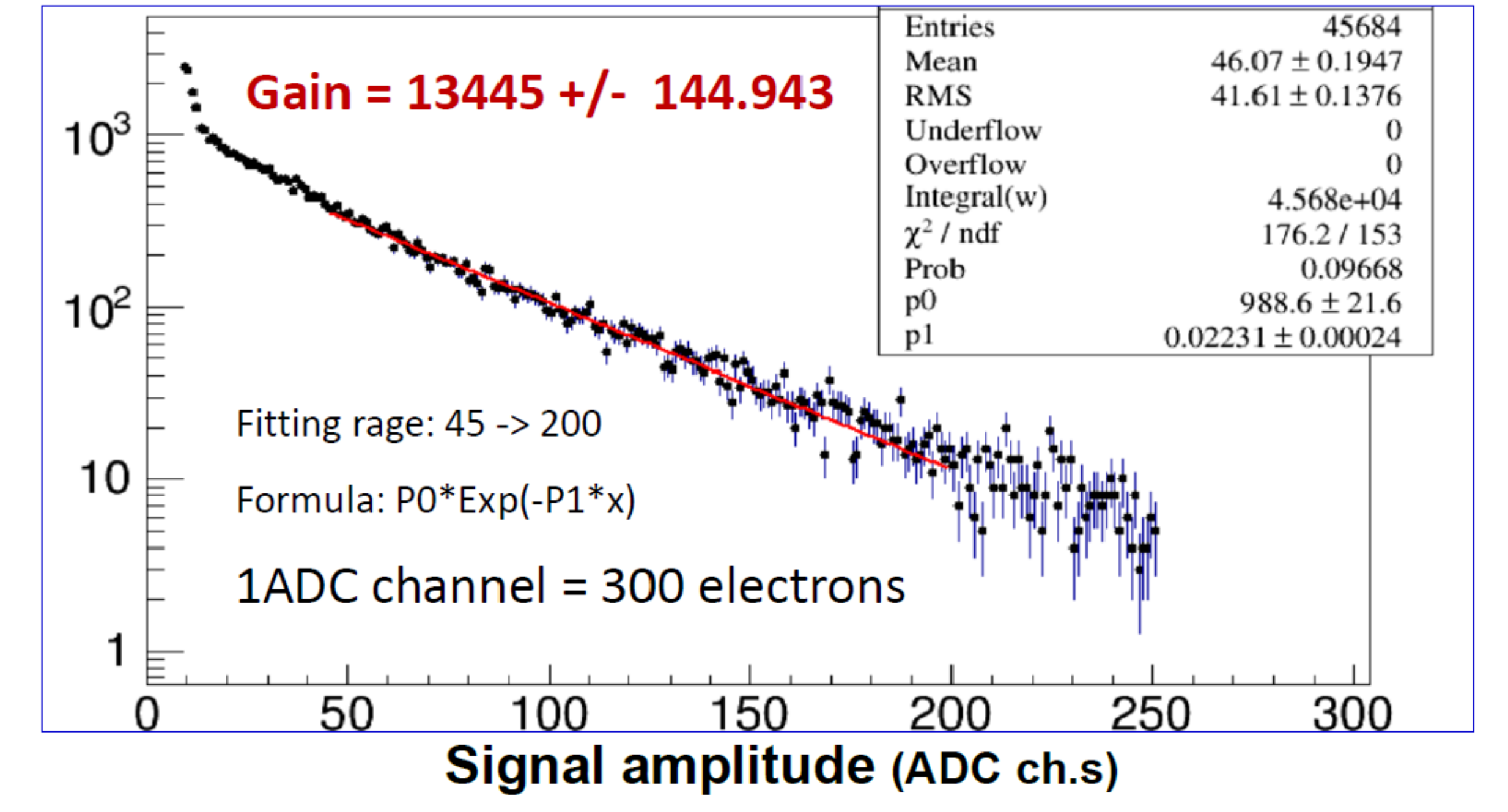}
\caption{Amplitude distribution for a sample of hits contributing 
to ring Cherenkov images.}
\label{fig:gain}
\end{figure}

The images generated in the photon detectors are clean and 
affected by very limited background (Fig.~\ref{fig:rings}).
The detector resolution in the measurement of the Cherenkov angle 
from single photoelectrons is  1.7-1.8~mrad r.m.s., fully
matching the expectation (Fig.~\ref{fig:resolution}). 
The amplitude 
spectrum of the photoelectron signals is expected 
to be exponential.
This is verified for pure photoelectron samples, 
obtained selecting 
hits contributing to ring images: the exponential behavior 
is present over more than two orders of
magnitude (Fig.~\ref{fig:gain}).
The detector gain is extracted from a fit of the spectrum 
and it ranges between 13k and 14k.  An electronic threshold 
of 3~times the noise level as measured pad by pad is applied
to each read-out channel.
The efficiency for single 
photoelectron detection is obtained from the gain and the 
threshold and it results higher than 80\%. The noise contributing 
to the ring images can be estimated from the spectrum deviation 
from a pure exponential function at small amplitude and it 
is at the 10\% level. A preliminary estimate of the number 
of detected photoelectrons 
per particle extrapolated to the saturation angle indicates 
11 photoelectrons.
\par
The high effective 
gain, the gain stability and the number of detected 
photoelectrons per ring
satisfy all the prerequisite requirements to ensure effective hadron
identification and stable performance
with the novel RICH-1 photon detectors.
\section{Future perspective}
The future Electron-Ion Collider (EIC)~\cite{eic} requires 
hadron identification at high momenta, a mission that can 
only be accomplished by RICH counters with an extended gas radiator.
The use of RICHes in the setup of collider experiments implies 
specific challenges. The radiator cannot be too extended to 
limit the overall apparatus size, imposing the need 
to detect more photoelectrons per radiator unit length. 
The photon detectors have to operate in presence of magnetic field.
A recent test-beam exercise has demonstrated the possibility 
to increase the number of detected photoelectrons by selecting 
the far UV range around 120~nm~\cite{windowless}. For this purpose, the 
RICH prototype has been operated window-less and CF$_4$ has been used at 
the same time as radiator gas and detector gas.
Therefore, we have started an R\&D program to 
match these specific requirements. It includes 
the exploratory study of a new option for the photoconverter: 
coating by hydrogenized nanodiamond powder~\cite{nd}.
\section{Conclusion}
The preliminary results obtained in  the characterization of 
novel MPGD-based photon detectors, in particular high effective 
gain, gain stability and number of photons per ring, indicate that they 
will fully accomplish the mission of increasing the 
stability and efficiency
of the photon detector system of COMPASS RICH-1.
They also represent a
technological achievement. In fact, for the first time in a 
running experiment, THGEMs are successfully used, 
single photon detection is accomplished by MPGDs, 
MPGDs are operated at gains larger than 10k.
\par
We have offered indications that MPGD-based photon detectors 
have a mission to accomplish also in the future, in particular in the hadron physics sector. 
\section*{Acknowledgments}
The authors are grateful to the colleagues of the 
COMPASS Collaboration for continuous support and encouragement.
\par
The use of the read-out system, 
originally designed and built for the MWPC with CsI photocathodes 
by the Munich and Saclay COMPASS groups, is a crucial ingredient 
for the successful performance of the MPGD-based photon 
detectors in COMPASS RICH.
\par
This work is partially supported by the 
H2020 project AIDA-2020, GA no. 654168.
J. Agarwala and Triloki are supported by ICTP TRIL fellowships.

\smallskip
\end{document}